\documentclass[a4j]{article}
\usepackage{amsmath,amssymb}
\usepackage{cite}
\usepackage[dvipdfm]{graphicx} 
\usepackage{float}
\usepackage{comment}
\usepackage[font=scriptsize]{caption}
\renewcommand{\figurename}{Fig.}
\date{}

\title{Generic temperature compensation of biological clocks by autonomous regulation of catalyst concentration}
\author{Tetsuhiro S. Hatakeyama\thanks{Department of Basic Science, Graduate School of Arts and Sciences, The University of Tokyo, 3-8-1 Komaba, Meguro-ku, Tokyo, 153-8902, Japan} \and Kunihiko Kaneko\footnotemark[1]}

\begin{document}

\maketitle

\begin{abstract}
Circadian clocks ubiquitous in life forms ranging bacteria to multi-cellular organisms, often exhibit intrinsic temperature compensation; the period of circadian oscillators is maintained constant over a range of physiological temperatures, despite the expected Arrhenius form for the reaction coefficient.
Observations have shown that the amplitude of the oscillation depends on the temperature but the period does not---this suggests that although not every reaction step is temperature independent, the total system comprising several reactions still exhibits compensation.
We present a general mechanism for such temperature compensation.
Consider a system with multiple activation energy barriers for reactions, with a common enzyme shared across several reaction steps with a higher activation energy.
These reaction steps rate-limit the cycle if the temperature is not high.
If the total abundance of the enzyme is limited, the amount of free enzyme available to catalyze a specific reaction decreases as more substrates bind to common enzyme.
We show that this change in free enzyme abundance compensate for the Arrhenius-type temperature dependence of the reaction coefficient.
Taking the example of circadian clocks with cyanobacterial proteins KaiABC consisting of several phosphorylation sites, we show that this temperature compensation mechanisms is indeed valid.
Specifically, if the activation energy for phosphorylation is larger than that for dephosphorylation, competition for KaiA shared among the phosphorylation reactions leads to temperature compensation.
Moreover, taking a simpler model, we demonstrate the generality of the proposed compensation mechanism, suggesting relevance not only to circadian clocks but to other (bio)chemical oscillators as well.
\end{abstract}

\section{Introduction}
The circadian clock is one of the most remarkable cyclic behaviors ubiquitous to the known forms of life, ranging from the unicellular to the multicellular level---including prokaryotes. Because of its importance, the underlying chemical reactions have been the subject academic interest for a long time and have recently been elucidated experimentally.
Circadian clocks have three important features: \\
1. They persist in the absence of external cues with an approximately 24-h period, which is rather long compared with most chemical reactions.\\
2. They can be reset by exposure to external stimuli such as changes in illumination (dark/light) or temperature.\\
3. The period of the circadian clock is robustly maintained across a range of physiological temperatures. (temperature compensation) \cite{Pittendrigh1954, Hastings1957} .

The emergence of cyclic behavior with a capacity for entrainment is theoretically understood as being result of the existence of a limit-cycle attractor in a class of dynamical systems described by chemical rate equations.
However, the phenomenon of temperature compensation is not yet fully understood.
Generally, the rate of chemical reactions depends strongly on the temperature.
Most biochemical reactions, in particular, have an energy barrier that must be outcome with the aid of enzymes, and thus the rate can be expected to follow the Arrhenius form.
Thus, the period of chemical or biochemical oscillators can be expected to strongly depend on the temperature \cite{Dutt1993}.
Thus, the ubiquitous temperature-compensation ability of biological circadian clocks, suggests that there may be some common mechanism(s) behind it.

Overall, there are two possibilities: one is that compensation exists at each elementary step, and the other is that compensation occurs at the system level---for the total set of enzymatic reactions.
Recently, it was found that the rates of some elementary reaction steps in circadian clocks depend only slightly on temperature \cite{Terauchi2007, Isojima2009}, suggesting that the activation energy barrier of some reactions is rather low.
Although some element-level compensation is important, it is difficult to imagine that every reaction step is fully temperature-compensated at the single-molecule level.
Indeed, if that were the case, temperature changes would not influence the oscillation of chemical concentrations at all.
However, even though the period of oscillation is generally insensitive to temperature changes, it is known that the amplitude changes indeed depend on temperature \cite{Liu1998, Majercak1999, Nakajima2005}.
Furthermore, circadian clocks are known to entrain to external temperature cycles, and they can be reset by temperature cues \cite{Liu1998, Merrow1999, Yoshida2009}.
Thus, although temperature changes can influence the oscillation, the period is still robust.
Hence, it is necessary to search for a general logic that underlies the temperature compensation phenomenon at the system level.
Indeed, several models have been proposed \cite{Ruoff1992, Leloup1997, Hong1997, Kurosawa2005, Hong2007}.
In most of the studies, several processes that are responsible for the period are considered, which cancel out the temperature dependence with each other.
With such balance mechanism, the temperature compensation is achieved.
These mechanisms, however, need fine-tuned set of parameters, or rather ad-hoc combination of processes for the cancellation.
Considering the ubiquity of temperature compensation, a generic and robust mechanism that dose not require tuning parameters is desirable.
Here we propose such a mechanism that has general validity for any biochemical oscillator consisting of several reaction processes catalyzed by enzymes.
The mechanism can be briefly outlined as follows:

Biochemical clocks comprise multiple processes, such as phosphorylation and dephosphorylation, with each generally having a different activation energy barrier $\Delta E$.
The rate of such reactions, then, is proportional to $A_f \exp(-\Delta E/T)$, where $A_f$ as the concentration of "free" enzyme available.
If the enzyme concentration were insensitive to temperature, the rate would just agree with the simple Arrhenius form, thus implying high temperature dependence.
Now, consider a situation where several substrates share the same enzyme.
At lower temperature, the reactions with higher activation energy will be slower and the substrates involved in these reactions will accumulate.
Then, because they share the same enzyme, competition for the enzyme will also increase.
Accordingly, the concentration of available (free) enzyme decreases, and when it reaches a level satisfying $A_f \exp(-\Delta E/T) \sim 1$, these enzymatic reactions will be highly suppressed.
The system spends most of its time under such conditions, which limits the rate.
Thus, the Arrhenius-type temperature dependence is compensated for by the concentration of available enzyme as  $A_f \exp(-\Delta E/T) \sim 1$.

Although this is a rather basic description, its validity may suggest that temperature compensation emerges in any general chemical oscillation consisting of steps, catalyzed by a common enzyme.
Here, we first study the validity of this enzyme-limited temperature compensation mechanism for the specific case of Kai protein clock model introduced by van Zon et al. \cite{VanZon2007} to explain the circadian clock of KaiABC proteins in cyanobacteria, which was discovered by Kondo and his colleagues \cite{Nakajima2005, Ishiura1998}
Indeed, in this system, the period of circadian oscillation of Kai proteins is temperature compensated \cite{Nakajima2005}.
Further, some of the elementary components in this system, specifically, KaiC's ATPase activity and KaiC's phosphatase activity, were suggested to depend only slightly on temperature \cite{Terauchi2007, Tomita2005}, but the origin of system-level temperature compensation has not yet been explained.
Here, we show numerically that system-level temperature compensation emerges from the differences in activation energy between phosphorylation and dephosphorylation and competition for KaiA as the enzyme that catalyzes KaiC phosphorylation.
Furthermore, we elucidate the conditions necessary for this temperature compensation to work.
Then, based on this analysis, we introduce a simpler model consisting of few catalytic reactions, to illustrate the above mechanism.
Possible relations between our results and reported experimental findings for circadian clocks are also discussed.

\section{Model}
The Kai-protein-based circadian clock, discovered by Kondo's group consists of KaiA, KaiB, KaiC proteins with ATP as an energy source \cite{Nakajima2005}.
KaiC has a hexameric structure with six monomers, each with two phosphorylation sites \cite{Kageyama2003}.
It has both self-kinase activity and self-phosphatase activity, but the self-phosphatase activity is usually stronger, and so it is spontaneously dephosphorylated \cite{Nishiwaki2000, Xu2003}.
KaiA, in a dimer form \cite{Kageyama2003}, attaches to KaiC and thus increases its kinase activity, leading to phosphorylation of KaiC \cite{Iwasaki2002, Xu2003}, while KaiB inhibits the activity of KaiA \cite{Williams2002, Xu2003}.
This phosphorylation/dephosphorylation process of the KaiABC proteins constitutes a circadian rhythm.
Here, we simplify the process, to focus on the temperature compensation of the period.
We reduce the two phosphorylation residues to just one, because abundance of singly phosphorylated KaiC is strongly correlated with that of doubly phosphorylated KaiC \cite{Rust2007} so that the phosphorylation of the two residues are equilibrated on a rather short time scale.
Next, we do not include KaiB explicitly in our model, because changes in the concentration of KaiB affect the period only slightly \cite{Nakajima2010}.
Note that although KaiB is necessary to generate circadian oscillations, the effect can be accounted for by introducing a parameter value for KaiA activity.
Here, we adopt a slightly simplified version of the model introduced by van Zon, et al. \cite{VanZon2007}(See also \cite{Yoda2007})

First, each KaiC monomer has two states---active and inactive. 
Second, allosterically regulated KaiC hexamers in the active state can be phosphorylated, whereas those in the inactive state can be dephosphorylated.
A phosphorylated KaiC monomer energetically prefers the inactive state, whereas a dephosphorylated KaiC has the opposite tendency.
Here the flip-flop transition between active and inactive states occurs only from the fully phosphorylated or fully dephosphorylated states, as assumed in the concerted MWC model \cite{Monod1965}.
No intermediate states are assumed.
Hence, the reaction process exhibits a cyclic structure as in Fig. 1.

\begin{figure}[H]
\begin{center}
　　　\includegraphics[clip]{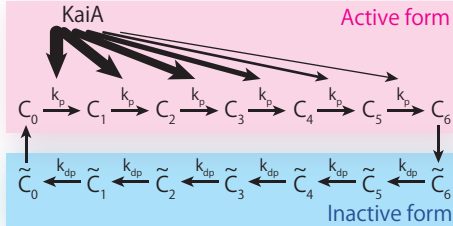}
　　　\caption{Simplified model of circadian clock involving KaiAC proteins based on van Zon et al. \cite{VanZon2007}
A KaiC hexamer, with six phosphorylation sites, can also take an active or inactive state.
In the active state, KaiC phosphorylation is catalyzed by KaiA at the rate $k_p$, while in the inactive state, dephosphorylation progresses without any enzymes at the rate $k_{dp}$.
Affinity between active KaiC and KaiA reduces as the number of phosphorylated sites of KaiC increases successively.
}
　　　\label{Fig1}
\end{center}
\end{figure}

Next, KaiA facilitates phosphorylation of active KaiC with an affinity that depends on the number of phosphorylated residues of each KaiC hexamer.
KaiCs with a smaller phosphorylation number have stronger affinity to KaiA and are phosphorylated faster.
This assumption is necessary for generating stable oscillations \cite{VanZon2007}.
Then, the reactions are given by

\begin{equation}
C_6\xrightarrow{f}\tilde{C}_6
\end{equation}
\begin{equation}
\tilde{C}_0 \xrightarrow{b} C_0 
\end{equation}
\begin{equation}
\tilde{C}_i \xrightarrow{k_{dp}} \tilde{C}_{i-1} 
\end{equation}
\begin{equation}
C_i + A \overset{k^{Af}}{\overrightarrow{\xleftarrow[k_i^{Ab}]{}}} AC_i \xrightarrow{k_p} C_{i+1} + A 
\end{equation}

Here $C_i$ and $\tilde{C}_i$ denote the concentrations of active KaiC and inactive KaiC, respectively, with $i$ phosphorylated sites; $A$ denotes the concentration of free KaiA dimer.
To study the temperature compensation of the period, we must also account for the temperature dependence of the reaction rate. 
Here, the rates of phosphorylation and dephosphorylation are governed by the Arrhenius equation.
Then the rate constants $k_{dp}$ and $k_p$ are as follows:

\begin{equation}
k_{dp} = c_{dp} \exp(-\beta E_{dp})
\end{equation}
\begin{equation}
k_p = c_p \exp(-\beta E_p)
\end{equation}\\
with inverse temperature $\beta$ ($\beta= 1/T$) by taking the unit of Boltzmann constant as unity.
We could include the temperature dependence of the rates $f$ and $b$ in a similar manner, but as the reaction between active and inactive states progresses faster and does not influence the period \cite{VanZon2007}, this dependence is neglected.
van Zon et al. demonstrated that temperature compensation occurs when the speed of phosphorylation and dephosphorylation is completely temperature-compensated at a level of  elementary reaction process.
However, it cannot explain why the amplitude of oscillation depends on temperature, or the entrainment to temperature cycle occurs.
The compensation mechanism at a system-level is wanted.
Here, the formation and dissociation of KaiAC complexes occur at much faster rates than other reactions and so are eliminated adiabatically.
Thus, the change in the concentration $A$ is given by

\begin{equation}
A_{total} = A + \sum_{i=0}^{5} \frac{A C_i}{K_i + A} \label{A}
\end{equation}\\
where $K_i (= k_i^{Ab} / k^{Af})$ are the dissociation constants.
Considering the increase in affinity for KaiA with the number of phosphorylated sites, we set $K_i = K_0 \alpha^i (\alpha > 1.0)$.
We adopted the deterministic rate equation given by the mass-action kinetics, and it is simulated by using the fourth-order Runge-Kutta method.

\section{Result}
\subsection{Oscillation at varied temperature}
As mentioned, a KaiC allosteric model was analyzed.
Specifically, we study the case in which the activation energy for phosphorylation, $E_p$, is larger than that of dephosphorylation, $E_{dp}$.
See the supplementary information for other cases.

\begin{figure}[H]
\begin{center}
　　　\includegraphics[clip,scale=0.5]{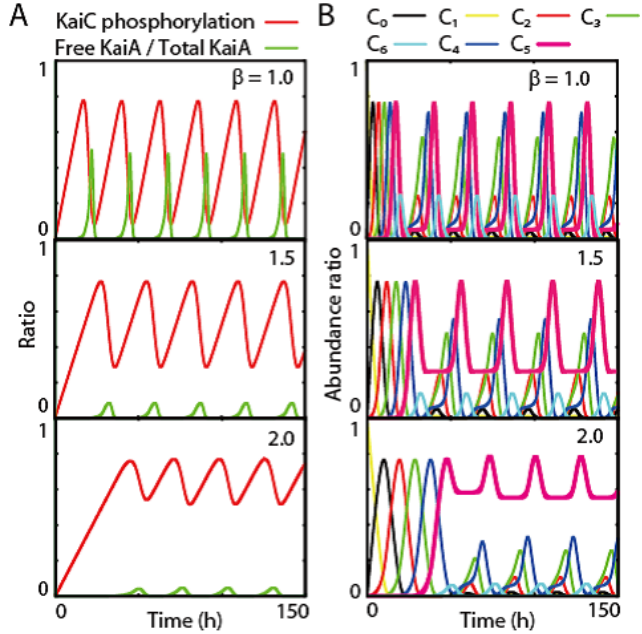}
　　　\caption{Oscillations in KaiC phosphorylation at various temperatures ($\beta = 1.0, 1.5, 2.0$).
(A) Red line indicates the time series of the mean phosphorylation level defined by $\Sigma_{i=0}^{6} i C_i / C_{total}$, whereas the green line indicates that of the fraction of free KaiA, $A / A_{total}$.
A decrease in temperature causes a decrease in the amplitude of the phosphorylation level.
(B) Time course of the abundance of each form of KaiC, $C_i$.
At low temperature, the basal amount of $C_5$ (magenta line) is remarkably high.
}
　　　\label{Fig2}
\end{center}
\end{figure}

For a certain range of parameter values, we found periodic oscillation in the KaiC phosphorylation level and free KaiA abundance, as shown in the time series in Fig. 2A.
The oscillation is described by a limit-cycle attractor, as represented in the orbit in a two-dimensional plane of KaiC phosphorylation level and free KaiA abundance.
As the temperature increases, the amplitude of the limit-cycle increases (Fig. 2A). 
Lowering the temperature causes a decrease in the maximum amount of free KaiA and increase in the minimum level of KaiC phosphorylation.
Further lowering it, however, results in the limit-cycle changing into a stable fixed point via Hopf bifurcation.
The timeseries of $C_i (i=0,1,2,..,6)$ and the temperature dependence are shown in Fig. 2B for $\beta$ = 1.0, 1.5, 2.0.
With a decrease in temperature, we see an increase in $C_5$, that is, the abundance of KaiC with five phosphorylated residues (5P-KaiC), which leads to an increase in the minimum KaiC phosophorylation level, as shown in Fig. 2B.
Note that there is a remarkable change in the time course of $C_5$ (the abundance of 5P-KaiC) at $\beta \simeq 1.2$ (Fig. 4A).
Below this inverse temperature, the minimum $C_5$ is close to zero.
At higher $\beta$ (i.e., lower temperature), however, $C_5$ never comes close to zero, and the minimum value increases with lowering of temperature. 

\subsection{Transition to temperature-compensated phase}
\begin{figure}[H]
\begin{center}
　　　\includegraphics[clip]{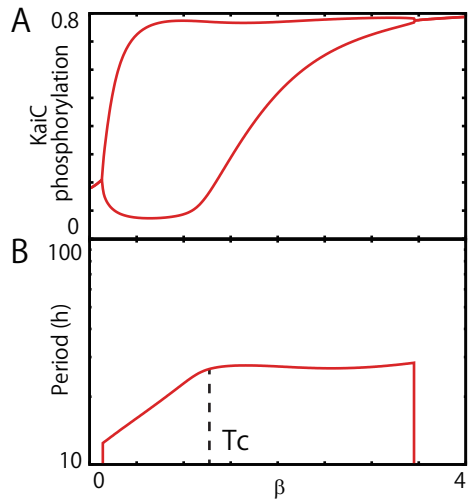}
　　　\caption{Dependence of amplitude and period of oscillation on the inverse temperature $\beta = 1/T$.
(A) Maximum and minimum values of mean phosphorylation level over a cycle.
The maximum value is nearly constant against temperature changes, whereas the minimum value increases with $\beta$ above $\beta_c = 1/ T_c \approx 1.2$, i.e., at temperatures below the characteristic temperature $T_c$.
The oscillation disappears via Hopf bifurcation at $\beta \approx 3.5$. 
(B) Period of the oscillation. 
At high temperature (low $\beta$), the period changes exponentially with $\beta$, whereas below $T_c$, the period is nearly constant against changes in temperature.
}
　　　\label{Fig3}
\end{center}
\end{figure}

\begin{figure}[H]
\begin{center}
　　　\includegraphics[clip]{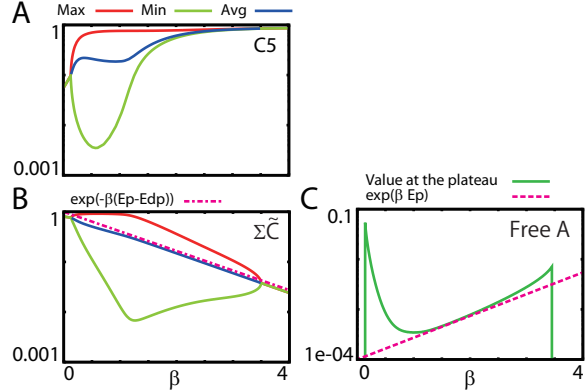}
　　　\caption{Normalized concentration of each component, $C_5 / C_{total}$ (A), $\Sigma \tilde{C} / C_{total}$ (B), $A / A_{total}$ (C) plotted against the inverse temperature $\beta$.
(A)(B) Maximum (red), minimum (green), and average (blue) over a cycle.
(A) Note that the maximum value of $C_5$ is nearly constant, wheras its minimum and average increase with $\beta$ beyond $\beta_c $ (i.e., at lower temperature).
(B) The average value of $\Sigma \tilde{C} / C_{total}$ is fitted well by the value of the unstable fixed point of the equation (magenta line), which is proportional to $\exp(-\beta(E_p-E_{dp})$.
(C) The concentration of free KaiA in the plateau region of $C_5$ is plotted, as estimated from the time where $C_3$ reaches a peak. 
This free KaiA concentration closely follows $\exp (\beta E_p)$ (magenta line), for $\beta>\beta_c$, i.e., below the characteristic temperature.
}
　　　\label{Fig4}
\end{center}
\end{figure}

The transition at $\beta \simeq \beta_{c} = 1/T_{c} \approx 1.2$ is also reflected in the temperature dependence of the period.
We plotted the period of oscillation as a function of (inverse) temperature, together with the maximum and minimum KaiC phosphorylation levels (see Fig. 3).
Above $T_{c}$, the temperature dependence of the period follows $\exp(\beta E_p)$, as can be naturally expected from a reaction process with a jump beyond the energy barrier. 
However, at lower temperature, the period is no longer prolonged exponentially and is nearly constant. 
Thus, the temperature compensation of the circadian period appears at lower temperature.
There is also a clear difference in the amplitude of oscillation below and above $T_c$.
At higher temperature (without temperature compensation), the amplitude of oscillation is almost constant over a large interval of temperatures.
However, at lower temperature (in the temperature-compensated phase), the amplitude decreases with lowering of temperature;
eventually, the oscillation disappears via Hopf bifurcation.
This decrease in amplitude is due to the increase in the minimum value of the KaiC phosphorylation level, caused by the increase in the minimum abundances of $C_5$.
The temperature dependences of the abundance of each KaiC and free KaiA also show distinct behaviors below and above $T_c$ (see Fig. 4, Supplementary Fig. 2).
The average and minimum abundances of $C_5$ increase remarkably with a decrease in temperature below $T_c$, whereas the maximum hardly changes with temperature (Fig. 4A). 
However, the amplitude of the oscillation of $C_i (0 \le i \le 4)$ has a peak at around $T_c$, and the minimum value increases as the temperature decreases.
The minimum as well as the plateau value of free KaiA increase when the temperature is decreased.
The minimum inactive KaiC abundance (independent of the residue number) and $C_6$ showed behaviors similar to those of $C_i (0 \le i \le 4)$, whereas the average followed $\exp(-\beta (E_p-E_{dp}))$ dependence throughout the temperature range.
Considering that the total amount of all KaiCs is conserved and the difference in activation energy, this dependence itself is rather natural.

\subsection{Proportion of free-energy of phosphorylation to that of dephosphorylation is critical for temperature compensation}
Thus, the transition in the oscillation and temperature compensation behavior at low temperature is the salient feature of the present system.
Next, we analyzed the conditions for $E_p$ and $E_{dp}$, the activation energies for phosphorylation and dephosphorylation, respectively, to elucidate the temperature compensation.
Supplementary Fig. 3 shows a plot for the region where temperature compensation appears in the parameter space of $E_p$ and $E_{dp}$.
From Supplementary Fig. 3, we observe that the temperature compensation appears in the regime $E_p \gtrsim 5 E_{dp}$.
The overall periodic behavior is determined mainly by $E_{dp} / E_{p}$, rather than the individual magnitudes.
When $0.2 < E_{dp} / E_{p} < 1$, there still exists a transition to a phase with weaker temperature dependence of period on lowering the temperature, but this effect is not sufficient to produce the temperature compensation (see also Fig. 5A).\\
(See supplementary informations for the case with $E_p < E_{dp}$.)

\subsection{Temperature compensation depends on KaiA and KaiC amounts}
\begin{figure}[H]
\begin{center}
　　　\includegraphics[clip]{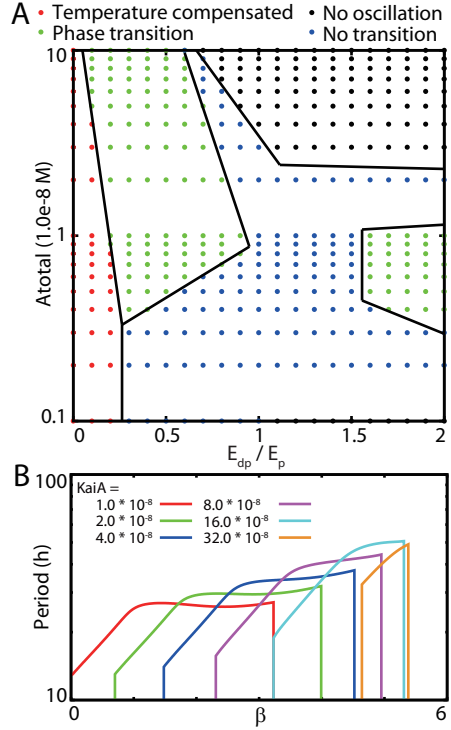}
　　　\caption{Influence of activation energy difference and the KaiA abundance on the temperature dependence of the period.
(A) Phase diagram of the temperature dependence of the cycle against the activation energy ratio ($E_{dp} / E_{p}$) and the abundance of total KaiA ($A_{total}$).
Red: temperature-compensated oscillation with $\partial \tau / \partial \beta \sim 0$ ($\tau$: the period) below the characteristic temperature.
Green: oscillation showing transition at the characteristic temperature.
Blue:  oscillation without transition, with a simple increase with temperature of the Arrhenius form.
Black: no oscillations at all.
As $| E_{dp} / E_p |$ increases from 0 to 1, we successively see temperature compensation, transition, and a disappearance of oscillation. (For the case with $E_p < E_{dp}$, see supplementary information). 
As $A_{total}$ increases, the width of the temperature compensation region decreases.
(B) Effect of KaiA increase on the period.
An increase in KaiA led to narrowing of the range of the periodic solution, and also the range of temperatures at which temperature compensation occurs.
}
　　　\label{Fig5}
\end{center}
\end{figure}

As already mentioned by von Zon et al., oscillation of KaiC abundance requires that the amount of KaiA is less than that of KaiC \cite{VanZon2007}.
An increase in KaiA abundance leads to decrease in the period, finally leading to no oscillations.
The range of temperatures where the oscillations exist narrows as KaiA abundance increases, and the oscillation disappears at higher temperatures.
Here, the transition temperature $T_c$ shifts to lower temperatures (see Fig. 5B).
Furthermore, the temperature compensation at $T < T_c$ is lost KaiA abundance is increased. 
Although the temperature dependence of the period is weaker at $T < T_c$, the dependence $\exp(\beta \Delta E')$ still exists with $\Delta E'$ smaller than $E_p$.
We plotted the range where temperature compensation occurs at $T < T_c$ in the two-dimensional plane with KaiA abundances and $E_{dp}/E_{p}$.
We see that low level of KaiA abundance is necessary for temperature compensation.

\subsection{Range of temperatures where systems is temperature-compensated is narrowed by a reduction in phosphorylation sites}
Since KaiC is a hexamer, we adopted six phosphorylation sites in our model.
However, to understand the biological significance of this number of sites, we examine models with a reduced number of phosphorylation sites (Supplementary Fig. 4).
We find that reducing the number of phosphorylation sites from hexamer to pentamer, and then to tetramer, narrows the temperature range where oscillations exist and temperature compensation occurs.
For the tetramer, temperature compensation is not observed even at $T < T_c$.

\subsection{KaiC phosphorylation cycle is entrained by temperature cycle}
\begin{figure}[H]
\begin{center}
　　　\includegraphics[clip]{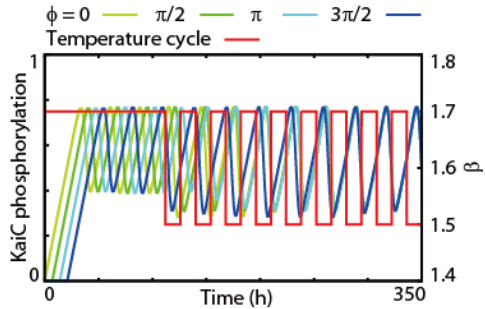}
　　　\caption{Entrainment of oscillation to externally applied thermal cycles.
Oscillations with various phases ($\phi = 0, \pi / 2, \pi, 3 \pi / 2$) are entrained by the temperature cycle between $\beta = 1.5$ and 1.7 within about 10 cycles.
}
　　　\label{Fig6}
\end{center}
\end{figure}

It has been experimentally it is suggested that KaiC's phosphorylation cycle is entrained to external temperature cycle \cite{Yoshida2009}.
To examine such entrainment, we cycle the temperature between $\beta = 1.5$ and $\beta = 1.7$ (i.e., within the region in the temperature compensation) periodically in time with a period close to that of the Kai system (27 h).
Within about 10 cycles, the phase of oscillation of the KaiC system is entrained with that of temperature, independently of the initial phase of oscillation.
Thus, entrainment to temperature cycle is achieved (see Fig. 6).

\section{Discussion}
Here we discuss how temperature compensation is achieved.
As presented in the Results section, two stages are necessary: transition in the temperature dependence of the period, and complete temperature compensation of the period at lower temperature.
As seen in the phase diagram (Fig. 5A), the former requires that $E_p$ is sufficiently larger than $E_{dp}$, which, as will be shown, means that the phosphorylation process is rate limited.
For the latter, the abundance of KaiA should be sufficiently small so that there is limited free KaiA that can be used for phosphorylation (i.e., competition for free KaiA).
As will be shown below, the abundance of free KaiA decreases as the temperature increases, which compensates for the increase in the rate constant of the reaction.

1. Phase transition

When there is a difference between the energy barriers for phosphorylation and dephosphorylation, the temperature dependence of the rate of each processes is different.
Roughly speaking, the time scale for the phosphorylation process changes in proportion to $A \exp(\beta E_p)$,  whereas the dephosphorylation has $\exp(\beta E_{dp})$, where $A$ is the concentration of free KaiA.
Thus, there exists a characteristic temperature at which the two rates are comparable; the rate-limited reaction switches at this temperature, where the phase transition to temperature dependence occurs.
Thus,

\begin{equation}
\beta_c \simeq \frac{\log|A|}{E_p - E_{dp}}
\end{equation}
where $A$, the concentration of free KaiA, is estimated from the steady-state solution of our model (see eq.(\ref{A})) to afford $A \simeq K_5 A_{total} / K_5 + C_5$ if $A_{total} \ll K_5$ and $A \simeq A_{total}$ if $A_{total} \gg K_5$.

Thus, a sufficient difference in activation energy is necessary for the phase transition-like behavior because the critical temperature will diverge as $E_p \simeq E_{dp}$.
If the difference is small, the critical temperature goes beyond the temperature for the onset of oscillation and the transition never occurs.
Moreover, if $A_{total}$ is too large or too small, the critical temperature is lower or higher than the range where the oscillation exists, and thus for both the cases, the phase transition disappears.
These estimates agree with the phase diagram in Fig. 5A.

2. Temperature compensation by self-adjustment of the KaiA concentration

When the temperature is lower than the transition temperature, the phosphorylation process takes more time, and $C_5$ is accumulated before dephosphorylation from $C_6$ progresses, as already discussed (see Fig.2B).
The increase in the abundance of active KaiC leads to competition for KaiA, and thus a decrease in free KaiA.
If the total KaiA abundance is limited, the system reaches a stage where phosphorylation almost stops.
This leads to the plateau in the time course of $C_5$, as observed in Fig. 2B.
This drastic slow down of the phosphorylation process occurs when $A$ is decreased to the level at ${K_p A \lesssim 1}$.
Thus, during the plateau in $C_5$, this approximate estimate gives,
$A \propto 1/ K_p \propto \exp(\beta E_p)$ ,
so that the temperature dependence of the phosphorylation rate $K_{p} A$ is compensated for by the decrease in $A$.
This plateau region is rate-limited in the circadian cycle, making the whole period independent of temperature.

In more detail, this compensation is also estimated as follows.
The abundances of the inactive forms of KaiC decrease with an increase in $C_5$.
Considering the differences in speed between phosphorylation and dephosphorylation, the total inactive KaiC abundance (at the fixed point) is estimated as

\begin{equation}
\Sigma \tilde{C}^* \propto \exp(-\beta (E_p - E_{dp})
\end{equation}\\
which is consistent with Fig. 4.
The flow from inactive KaiC is thus estimated by

\begin{equation}
k_{dp} \Sigma \tilde{C}^* \propto \exp(-\beta E_p).
\end{equation}

This flow starts the phosphorylation processes from $C_0$, but is slowed down at some residue number $m$.
In the present model, this slow down starts at $m \approx 2 \sim 3$ due to the paucity of free KaiA.
Following the above estimate of flow, the maximum $C_m$ is estimated to be proportional to $\exp(-\beta E_p)$.
Now from eq.(\ref{A}) and $K_m \ll K_i$ $(i > m)$ and $C_j \ll C_m$ $(j < m)$ at this time, the abundance of free KaiA can be estimated approximately as $A \simeq A_{total} - (A C_m)/(K_m + A)$
When $A_{total}$ is small enough, the minimum free KaiA abundance is smaller than $K_m$,thus $A \simeq A_{total} - (A C_m)/(K_m)$.

\begin{equation}
A \simeq \frac{A_{total}}{1+C_m/K_m} \simeq \frac{A_{total} K_m}{C_m} \propto \exp(\beta E_p).
\end{equation}

Therefore, when $A_{total}$ is small enough and $E_p$ is sufficiently greater than $E_{dp}$, the time scale of the phosphorylation process is $k_p A$, and the period of the cycle is temperature-compensated.
Indeed, as shown in Supplementary Fig. 2, the maximum $C_1$ (and $C_2$) values show $\exp(-\beta E_p)$ dependence, whereas free KaiA shows approximately $\exp(\beta E_p)$ dependence when the phosphorylation is slowed down.

In essence, the temperature compensation mechanism requires two properties: \\
difference in activation energy between phosphorylation and dephosphorylation processes, and a limited abundance of enzyme KaiA.
The former is essential to the phase transition, and the latter to the compensation of the Arrhenius-type by the temperature dependence of the individual reactions.
As long as these conditions are met, the period is temperature-compensated at low temperature, without the need for fine tuning of the parameters of the system.

These two properties generally appear if there are two types of processes with different activation energy, with one type catalyzed by a common enzyme and the other without catalyzation. If the enzyme abundance is limited, the competition for the enzyme will lead to temperature compensation in cyclic reaction systems.
In particular, if the main component of the chemical reactions has allosteric structure like KaiC, the competition for enzymes will occur naturally.

\begin{figure}[H]
\begin{center}
　　　\includegraphics[clip]{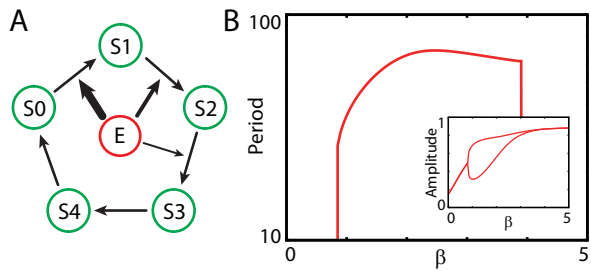}
　　　\caption{Temperature compensation of period in a simpler model with enzyme competition.
(A) Scheme of the simpler model.
The substrate has five forms, and reactions $S_0 \rightarrow S_1, S_1 \rightarrow S_2, S_2 \rightarrow S_3, S_3 \rightarrow S_4$ are catalyzed by the same enzyme $E$, but the other reactions are not.
The affinity between $S_i$ and $E$ weaken as $i$ increases.
(B) Maximum and minimum values of $S_2$ over a cycle and (C) the period, plotted against the inverse temperature $\beta$.
The oscillation is temperature-compensated over a specific range of temperatures, although the transition is not as sharp as in the Kai model.
}
　　　\label{Fig7}
\end{center}
\end{figure}

As an example, we consider the cyclic process shown in Fig. 7, where a cyclic change in five substrates occurs.
Three processes---$S_0\rightarrow S_1 \rightarrow S_2 \rightarrow S_3$---have a high activation energy barrier, and are catalyzed by the common enzyme $E$. 
In this case, the period of the oscillation in the concentration of each substrate is temperature-compensated at low temperature. 
(If the number of reactions with the common enzyme is decreased to two, compensation still appears, but it is weaker, as shown in Supplementary Fig. 5). 
This example demonstrates the generality of the present mechanism, and opens the possibility of applications to temperature compensation in other biochemical oscillations as well.

The regulation of reaction process by autonomous changes in enzyme concentration, as adopted here, was previously pointed out by Awazu and Kaneko, who reported that relaxation to equilibrium slowes down when the concentrations of substrate and enzymes are negatively correlated \cite{Awazu2009}.
Excess substrate hinders the enzymatic reaction, leading to a plateau in relaxation dynamics.
In the present model, the total concentration of KaiA as an enzyme is a conserved quantity but the fraction of free KaiA available is reduced when there is abundant active KaiC and so the speed of phosphorylation slows down dramatically.
Such regulation of reaction speed by limitation of catalysts should be important, not only for the temperature compensation of circadian clocks in general, but also for other biological processes \cite{Bullock1955}. 

Now we briefly discuss the relevance of the present results to experiments on the Kai protein circadian system.
In our model, it is assumed that KaiC's affinity to KaiA depends on the phosphorylation levels.
This assumption is necessary not only for the emergence of temperature compensation but also for the existence of the oscillation itself.
Recent reports that KaiC phosphorylation induces conformational changes may account for such changes in the binding affinity.

In the actual circadian clock comprising Kai proteins, KaiB is also involved besides KaiA and KaiC.
However, as mentioned, an increase in KaiB abundance has minor influence on the clock \cite{Nakajima2010}.
Earlier theoretical study suggested that KaiB binds to KaiAC complexes strongly and restricts the concentration of free KaiA  \cite{VanZon2007}.
Thus, the inclusion of KaiB is expected to not alter the present temperature compensation mechanism, but to facilitate it by strengthening the limitation of enzyme availability.
Note that previously reported temperature compensation at element-level---KaiC's ATPase activity \cite{Terauchi2007} and auto-dephosphorylation activity \cite{Tomita2005} is also relevant to  the system-level compensation in our model.
The former is relevant to achieve fast equilibration between KaiA and KaiA complex used for adiabatic elimination in eq.(\ref{A}), and the latter to provide $E_{dp}\ll E_p$.

It is known that the amplitude of KaiC oscillation decreases as the temperature is lowered \cite{Nakajima2005}, which agrees well with our results (Fig. 2A).
Indeed, as described already, this decrease in amplitude is tightly coupled with the temperature compensation mechanism.
It is also interesting to note that in many circadian clocks, the period does not depend on the temperature, although the amplitude decreases with temperature \cite{Nakajima2005}.
If every elementary step of the circadian clock were temperature compensated at the single-molecule level, this temperature dependence of the amplitude would not be possible.
Furthermore, the entrainment of the KaiC oscillation to imposed temperature cycles, as observed in a recent experiment would also not be possible \cite{Yoshida2009}.
In our study, the compensation is based on the different temperature dependences of the phosphorylation and dephosphorylation processes; the entrainment is a result of this difference.

Our mechanism for temperature compensation depends on paucity of KaiA---increasing the KaiA concentration leads to loss of the compensation.
We expect that this prediction will be directly confirmed in a future experiment.

\section*{Acknowledgments}
The authors would like to thank A. Awazu, H. Iwasaki, Y. Murayama, H. R. Ueda, T. Yomo for useful discussion.

\newpage

\setcounter{figure}{0}
\renewcommand{\figurename}{Supplementary Fig}

\begin{figure}[H]
\begin{center}
　　　\includegraphics[clip]{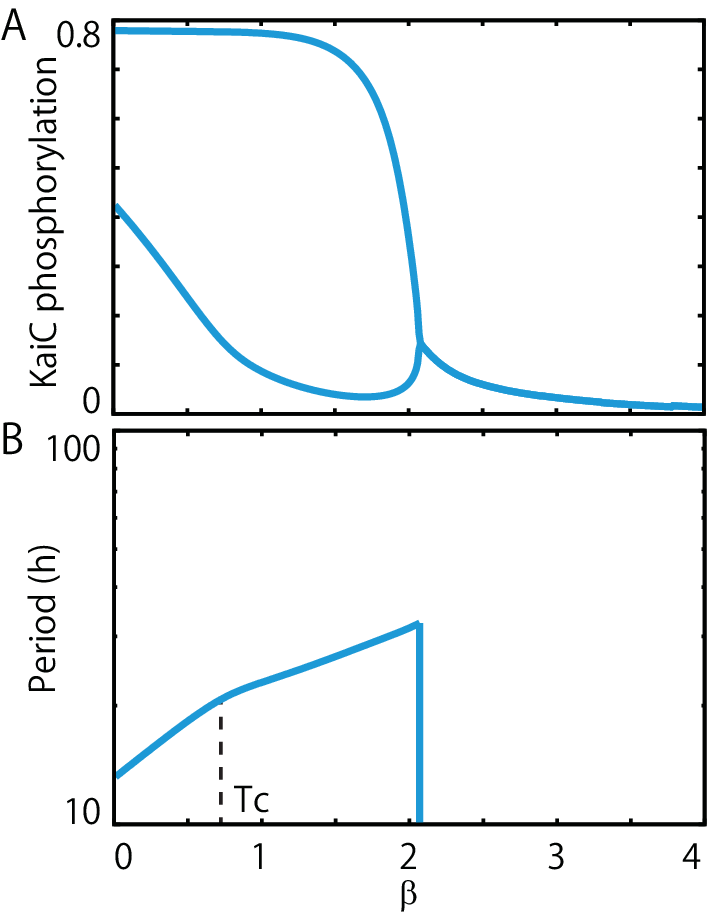}
　　　\caption{Temperature dependence of the amplitude and period of oscillation for $E_p < E_{dp}$ ($E_p = 0.1$, $E_{dp} = 1.0$).
(A) Maximum and minimum mean phosphorylation levels over a cycle.
The maximum value is nearly constant against temperature changes.
The minimum value increases with $\beta$ for $\beta >\beta_c = 1/ T_c \approx 0.7$, and is nearly constant below $\beta_c$.
The oscillation disappears at $\beta \approx 2.0$.
(B) Period of oscillation.
The period changes with the Arrhenius form, i.e., in an exponential manner over the whole range of temperature, while the slope changes below and above $T_c$.
Below $T_c$ (beyond $\beta_c$), the temperature dependence of the period is about $\exp (\beta E_dp)$, and is $\exp(\beta \Delta E')$ with $\Delta E'$ larger than $E_{dp}$ at $T > T_c$ ($\beta<\beta_c$).
}
　　　\label{SFig2}
\end{center}
\end{figure}

\begin{figure}[H]
\begin{center}
　　　\includegraphics[clip]{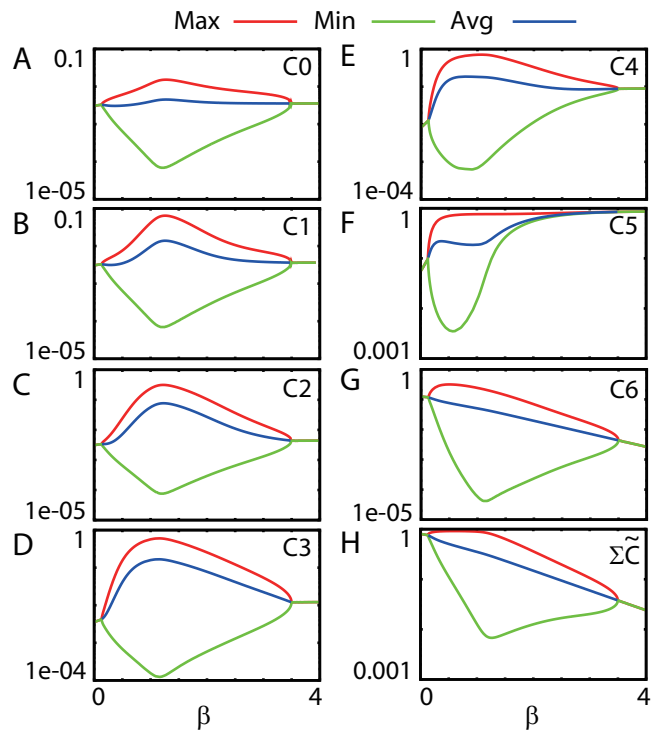}
　　　\caption{Concentration of each KaiC forms.
Red line : maximum value, green line : minimum value, blue line : average value over a cycle.
(A) $C_0$, (B) $C_1$, (C) $C_2$, (D) $C_3$, (E) $C_4$, (F) $C_5$, (G) $C_6$, (H) $\Sigma \tilde{C}$.
}
　　　\label{SFig3}
\end{center}
\end{figure}

\begin{figure}[H]
\begin{center}
　　　\includegraphics[clip]{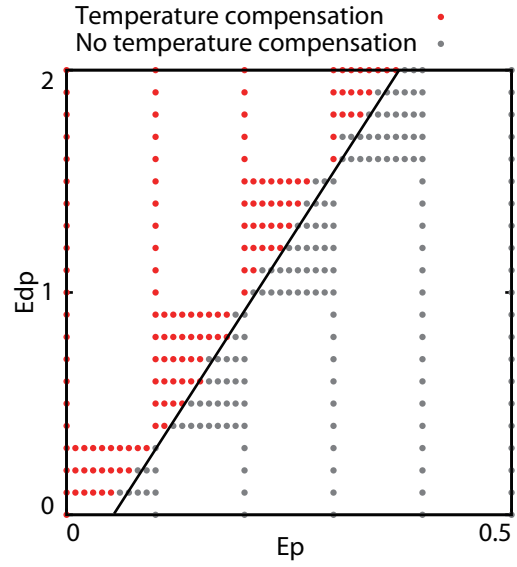}
　　　\caption{Phase diagram showing temperature dependence of the period for various $E_p$ and $E_{dp}$, while $A_{total}$ is fixed at $ 1.2 \times 10^{-8}$.
Red: temperature-compensated oscillation satisfying $\partial \tau / \partial \beta \leq 0$ ($\tau$ : the period) below the characteristic temperature.
Gray: oscillation without temperature compensation.
When $0.2 < E_{dp} / E_{p} < 1$, the oscillation is temperature-compensated at temperatures below the characteristic temperature.
}
　　　\label{SFig4}
\end{center}
\end{figure}

\begin{figure}[H]
\begin{center}
　　　\includegraphics[clip]{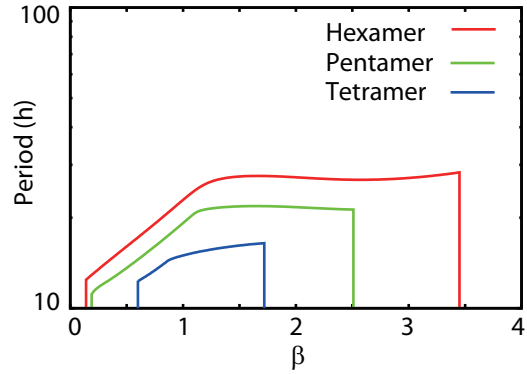}
　　　\caption{Temperature dependence of the period of reduced Kai models.
The hexamer is the original model shown in Fig. 1 containing $C_0 \sim C_6$, while a pentamer and tetramer are the reduced models containing $C_0 \sim C_5$ and $C_0 \sim C_4$, respectively. 
The same parameter values are used as in the original model.
Models with fewer phosphorylation sites ($i \leq 3$) cannot generate any oscillation.
In the reduced models, the temperature ranges where oscillations occur and where temperature-compensated oscillations occur are both narrower, and the temperature compensation in the tetramer model is not perfect.
}
　　　\label{SFig5}
\end{center}
\end{figure}

\begin{figure}[H]
\begin{center}
　　　\includegraphics[clip]{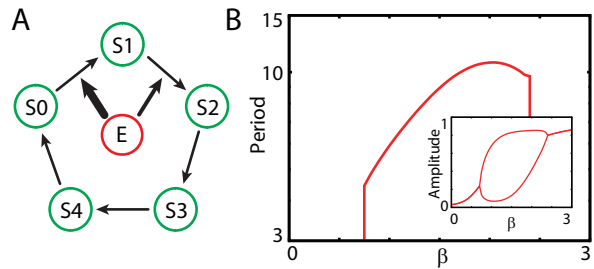}
　　　\caption{Temperature compensation in a simpler model with just two catalytic reactions sharing the same enzyme.
(A) Scheme of the model.
The reactions between $S_0 \rightarrow S_1, S_1 \rightarrow S_2, S_2
\rightarrow S_3$ are catalyzed by the same enzyme $E$, and other reactions
are not.
The affinity between $S_0$ and $E$ is higher than that between $S_1$ and $E$.
 (B) the period, plotted as a function of the inverse temperature $\beta$.
The range with temperature compensation is narrower, and the compensation is not perfect.
(Inset: Maximum and minimum values of $S_1$ over a cycle).
}
　　　\label{SFig6}
\end{center}
\end{figure}

\end{document}